\documentclass[twocolumn,pre,aps,showpacs]{revtex4}
\usepackage{graphicx}
\usepackage[T1]{fontenc}
\usepackage{times}

\begin{document}

\title{Correlated disordered interactions on Potts models}

\author{P. T. Muzy}
\email{ptmuzy@uol.com.br}
\author{A. P. Vieira}
\email{apvieira@if.usp.br}
\author{S. R. Salinas}
\email{ssalinas@if.usp.br}
\affiliation{Instituto de F\'{\i}sica, Universidade de S\~{a}o Paulo\\
 Caixa Postal 66318\\
 05315-970 S\~{a}o Paulo, SP, Brazil}
\date{\today}

\begin{abstract}
Using a weak-disorder scheme and real-space renormalization-group techniques,
we obtain analytical
results for the critical behavior of various \( q \)-state Potts models with
correlated disordered exchange interactions along \( d_{1} \) of \( d \) spatial
dimensions on hierarchical (Migdal-Kadanoff)
lattices. Our results indicate qualitative differences
between the cases \( d-d_{1}=1 \) (for which we find nonphysical random fixed
points, suggesting the existence of nonperturbative fixed distributions) and
\( d-d_{1}>1 \) (for which we do find acceptable
\textit{perturbartive} random fixed
points), in agreement
with previous numerical calculations by Andelman and Aharony. We also rederive
a criterion for relevance of correlated disorder, which generalizes the usual
Harris criterion.
\end{abstract}
\maketitle
\section{Introduction}

The effects of disorder on the critical properties of statistical models have
been the subject of much work in the last decades. In the context of random
interactions, Harris \cite{Harris} derived a heuristic criterion to gauge the
relevance of uncorrelated disorder to the critical behavior, which is predicted
to remain unchanged if the specific-heat exponent \( \alpha  \) of the underlying
pure system
is negative. If \( \alpha >0 \), disorder becomes relevant and, in the language
of the renormalization group (RG), one expects a flow to a new fixed point (characterized
by a nonzero-width fixed distribution of the random variables).

It later became clear that the Harris criterion must be generalized in a number
of situations \cite{Lubensky,ddy,andelman,Muk,efrat2001}, since \( \alpha  \)
is not always identifiable with \( \phi  \), the crossover exponent of the
width of the distribution of the disorder variables. In particular, random variables
correlated along \( d_{1} \) of the \( d \) spatial dimensions give rise to
the scaling relation \cite{Lubensky,andelman}
\begin{equation}
\label{eqlub}
\phi =\alpha +d_{1}\nu ,
\end{equation}
where \( \nu  \) is the correlation-length exponent of the pure system. Using
a real-space RG approach based on numerical calculations \cite{berker}, Andelman
and Aharony \cite{andelman} investigated various \( q \)-state Potts models
with random exchange constants, finding qualitative differences between the
cases \( d-d_{1}>1 \) (which yields finite-temperature fixed distributions)
and \( d-d_{1}=1 \) (which embodies the McCoy-Wu model \cite{McCoy and Wu}
and yields an ``infinite-disorder'' zero-temperature fixed point).
An intuitive illustration of the special role of the $d-d_{1}=1$ case is that,
for any infinitesimal concentration of zero bonds
(with a suitable assignment of the random interactions), the system would break
into non-interacting $(d-1)$-dimensional structures, and the RG flows would be
redirected to the pure fixed point of the corresponding system in $d-1$ dimensions.

In the present paper, we use a (perturbative) weak-disorder
\cite{Derrida and Gardner,Muzy and Salinas}
real-space RG scheme to analyze the critical behavior of \( q \)-state Potts
models with correlated disordered exchange interactions on various
hierarchical lattices,
whose exact recursion relations are equivalent to those produced by
Migdal-Kadanoff
approximations for Bravais lattices.
Using this weak-disorder scheme, we obtain
analytical results by truncating the recursion relations for the moments of
the disorder distribution (which are supposed to remain sufficiently small
under the RG iterations).
All calculations are performed in the vicinity of $\phi =0$, in a
region where disorder is relevant.
Depending on the diference between the
dimensionality of the system ($d$) and the number of dimensions
in which disorder is correlated ($d_{1}$),
we distinguish two possibilities:
(i) For $d-d_{1}=1$, the weak-disorder scheme
produces a nonphysical fixed-point probability distribution, characterized by
a negative variance, which suggests the existence of a nonperturbative
(``infinite-disorder'')
fixed-point; (ii) For $d-d_{1}>1$, the scheme yields a physically
acceptable perturbative fixed-point distribution. Although obtained by an
alternative approach, the main results of this paper are in agreement with
the numerical findings of Andelman and Aharony \cite{andelman}.

The outline of the paper is as follows.
We first rederive
Eq. (\ref{eqlub}), and obtain a criterion for relevance of correlated disorder
involving the number of independent random variables in the unit cell of the
lattice and the first derivative of the recursion relations at the pure fixed
point. This is done in Sec. \ref{criterion}.
In Sec. \ref{secp1}, we consider
\( q \)-state Potts models on various hierarchical lattices with $d-d_{1}=1$.
Using a weak-disorder scheme, we obtain a new (random) fixed point for \( q \)
larger than a characteristic value \( q_{0} \), where disorder becomes relevant.
As in a previous publication \cite{Muzy and Salinas}, this fixed point is located
in a nonphysical region of the parameter space, suggesting that a nonperturbative
fixed point must be present. In Sec. \ref{secp2} we
study a similar problem with \( d_{1}=1 \) and \( d=3 \). In this case
we obtain a physically acceptable, finite-disorder fixed point, for \( q>q_{0} \),
as in the fully disordered model studied by Derrida and Gardner
\cite{Derrida and Gardner}
(although in our case the usual Harris criterion is not satisfied). In Sec.
\ref{secising}, we consider
an Ising model (\( q=2 \)) on a diamond lattice with \( b=2 \) bonds and \( l \)
branches (where \( l \), instead of \( q \), is the control parameter), which
constitutes another example of a \( d-d_{1}=1 \) system. As
in Sec. \ref{secp1}, weak disorder again predicts a nonphysical random fixed
point. In the final section we give some conclusions.

\section{Criterion for relevance of correlated disorder}
\label{criterion}
Following Andelman and Aharony \cite{andelman}, we consider a \( d \)-dimensional
bond-disordered model in which the disorder variables are correlated along \( d_{1} \)
spatial directions. We assume that, under renormalization with a length rescaling
factor \( b \), the model satisfies a recursion relation \( R\left( x_{1},x_{2},\ldots ,x_{n}\right)  \),
connecting \( n=b^{d-d_{1}} \) independent (and identically distributed) random variables
to a renormalized variable \( x^{\prime } \).
(In this paper, these variables are related to reduced exchange couplings.)
Defining the deviations \( \varepsilon _{i}\equiv x_{i}-x_{c} \),
where \( x_{c}=R(x_{c},x_{c},\ldots ,x_{c}) \) is the critical fixed point
of the pure system, we expand \( R \) in a Taylor series about \( x_{c} \)
to write

\begin{widetext}

\begin{equation}
\varepsilon ^{\prime }\equiv x^{\prime }-x_{c}=\sum ^{n}_{i=1}\left. \frac{\partial R}{\partial x_{i}}\right| _{x_{c}}\varepsilon _{i}+\frac{1}{2}\sum ^{n}_{i,j=1}\left. \frac{\partial ^{2}R}{\partial x_{i}\partial x_{j}}\right| _{x_{c}}\varepsilon _{i}\varepsilon _{j}+\cdots ,
\end{equation}

\begin{equation}
\left. \varepsilon ^{\prime }\right. ^{2} = \sum ^{n}_{i,j=1}\left.
\frac{\partial R}{\partial x_{i}}\right| _{x_{c}}\left. \frac{\partial R}
{\partial x_{j}}\right| _{x_{c}}\varepsilon _{i}\varepsilon _{j}
+\sum ^{n}_{i,j,k=1}\left. \frac{\partial R}{\partial x_{i}}\right|
_{x_{c}}\left. \frac{\partial ^{2}R}{\partial x_{j}\partial x_{k}}\right|
_{x_{c}}\varepsilon _{i}\varepsilon _{j}\varepsilon _{k}+\cdots ,
\end{equation}
and similarly for the higher powers of \( \varepsilon ^{\prime } \). Averaging
over the random variables we get
\begin{equation}
\left\langle \varepsilon ^{\prime }\right\rangle =\sum ^{n}_{i=1}\left. \frac{\partial R}{\partial x_{i}}\right| _{x_{c}}\left\langle \varepsilon \right\rangle +\frac{1}{2}\sum _{i=1}^{n}\left. \frac{\partial ^{2}R}{\partial x_{i}^{2}}\right| _{x_{c}}\left\langle \varepsilon ^{2}\right\rangle +\sum _{i\neq j}\left. \frac{\partial ^{2}R}{\partial x_{i}\partial x_{j}}\right| _{x_{c}}\left\langle \varepsilon \right\rangle ^{2}+\cdots ,
\end{equation}
\begin{equation}
\left\langle \left. \varepsilon ^{\prime }\right. ^{2}\right\rangle =\sum ^{n}_{i=1}\left( \left. \frac{\partial R}{\partial x_{i}}\right| _{x_{c}}\right) ^{2}\left\langle \varepsilon ^{2}\right\rangle +\sum _{i\neq j}\left. \frac{\partial R}{\partial x_{i}}\right| _{x_{c}}\left. \frac{\partial R}{\partial x_{j}}\right| _{x_{c}}\left\langle \varepsilon \right\rangle ^{2}+\cdots ,
\end{equation}

\end{widetext}

\noindent and corresponding expressions for the higher moments of the deviations. Since
\( \left\langle \varepsilon \right\rangle  \) is a measure of the distance
to the fixed point, it plays the role of temperature. On the other hand, \( \left\langle \varepsilon ^{2}\right\rangle  \)
is a measure of the strength of disorder.

The critical behavior of the model is related to the eigenvalues of the matrix
\begin{equation}
M_{rs}=\frac{\partial \left\langle \left. \varepsilon ^{\prime }\right. ^{r}\right\rangle }{\partial \left\langle \varepsilon ^{s}\right\rangle },
\end{equation}
evaluated at the fixed point. It is clear that the set of recursion relations
for the moments of the deviations always has a pure fixed point \( \left\langle \varepsilon \right\rangle =\left\langle \varepsilon ^{2}\right\rangle =\cdots =0 \).
At that point, it can be shown \cite{Efrat} that \( M_{rs} \) is a triangular
matrix, and that its two largest eigenvalues are given by
\begin{equation}
\label{eql1}
\Lambda _{1}=\left. \frac{\partial \left\langle \varepsilon ^{\prime }\right\rangle }{\partial \left\langle \varepsilon \right\rangle }\right| _{\mathrm{pure}}=\sum ^{n}_{i=1}\left. \frac{\partial R}{\partial x_{i}}\right| _{x_{c}}
\end{equation}
and
\begin{equation}
\label{eql2}
\Lambda _{2}=\left. \frac{\partial \left\langle \left. \varepsilon ^{\prime }\right. ^{2}\right\rangle }{\partial \left\langle \varepsilon ^{2}\right\rangle }\right| _{\mathrm{pure}}=\sum ^{n}_{i=1}\left( \left. \frac{\partial R}{\partial x_{i}}\right| _{x_{c}}\right) ^{2}.
\end{equation}
Assuming that, for all \( i \) and \( j \),
\begin{equation}
\label{eqw}
\left. \frac{\partial R}{\partial x_{i}}\right| _{x_{c}}=\left. \frac{\partial R}{\partial x_{j}}\right| _{x_{c}}\equiv w,
\end{equation}
and invoking the usual scaling hypotheses
\begin{equation}
\Lambda _{1}=b^{y_{t}}\qquad \textrm{and}\qquad \Lambda _{2}=\Lambda ^{\phi }_{1}=b^{\phi y_{t}},
\end{equation}
which define the thermal exponent \( y_{t} \) and the crossover exponent \( \phi  \),
we get
\begin{equation}
\phi y_{t}=2y_{t}-\left( d-d_{1}\right).
\end{equation}
Then, using the hyperscaling relation
\begin{equation}
\alpha =2-\frac{d}{y_{t}}=2-\frac{d\ln b}{\ln (nw)},
\end{equation}
we obtain
\begin{equation}
\phi =\alpha +\frac{d_{1}}{y_{t}}=\frac{d-d_{1}}{d}\alpha +2\frac{d_{1}}{d},
\end{equation}
which clearly shows that the Harris criterion (\( \phi =\alpha >0 \)) is not
satisfied in the presence of correlated disorder. As \( 1/y_{t} \) is usually
identified with the correlation-length exponent \( \nu  \), this last result
is equivalent to Eq. (\ref{eqlub}). It also shows that, for \( d_{1}>0 \), the
crossover expoent is larger than \( \alpha  \), which indicates that correlated
disorder induces stronger (geometrical) fluctuations than uncorrelated disorder.

The general criterion for relevance of disorder is \( \phi >0 \), that is,
\begin{equation}
\label{eqcrit}
\alpha >-2\frac{d_{1}}{d-d_{1}}.
\end{equation}
From Eqs. (\ref{eql1})-(\ref{eqw}), this is equivalent to
\begin{equation}
\label{eq11}
nw^{2}>1.
\end{equation}
This last result was also derived in a different context by Mukherji and Bhattacharjee
\cite{Muk} and generalizes a criterion pointed out by Derrida \emph{et al.} \cite{ddy}.

In the case of the fully disordered system analyzed by Derrida and Gardner \cite{Derrida and Gardner},
for which \( d_{1}=0 \), the requirement in Eq. (\ref{eqcrit}) turns out to
be equivalent to the usual form of the Harris criterion (\( \alpha >0 \)).

\section{Potts models with correlated disorder: \lowercase{$d-d_{1}=1$} case}
\label{secp1}
\begin{figure}
\includegraphics[width=5.0cm]{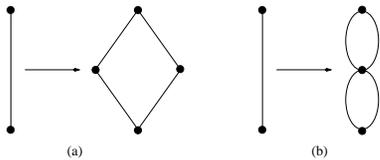}
\caption{(a) The diamond hierarchical lattice (with $b=2$ and $l=2$).
(b) The necklace hierarchical lattice (with $b=2$ and $l=2$).
\label{fig1}}
\end{figure}

The successive generations of a hierarchical lattice are obtained by replacing
an existing bond in the previous generation by a unit cell of new bonds in the
next generation. In Fig. \ref{fig1}(a), we show the first two stages of the construction
of the simple diamond lattice (with \( b=2 \) bonds and \( l=2 \) branches).
The necklace hierarchical lattice, with \( b=2 \) bonds and \( l=2 \) branches,
is illustrated in Fig. \ref{fig1}(b).

\begin{figure}
\includegraphics[width=6.0cm]{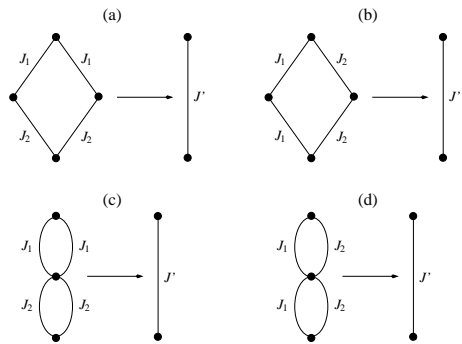}
\caption{Correlated distribution of random interactions on diamond and necklace
hierarchical lattices.
\label{fig2}}
\end{figure}

We now consider a \( q \)-state Potts model, given by the Hamiltonian
\begin{equation}
\label{eq01}
{\mathcal{H}}_{P}=-\sum _{(i,j)}J_{ij}\delta _{\sigma _{i},\sigma _{j}},
\end{equation}
 where the sum is over nearest-neighbor sites on a hierarchical lattice, the
spin variables \( \sigma _{i} \) assume \( q \) values, \( \delta  \) is
the Kronecker symbol, and \( \left\{ J_{ij}>0\right\}  \) is a set of independent
and identically distributed random variables. Instead of considering a fully
disordered arrangement of interactions, we look at correlated disorder, either
along layers (see Figs. \ref{fig2}(a) and \ref{fig2}(c)) or along branches
(see Figs. \ref{fig2}(b) and \ref{fig2}(d)) of the hierarchical structure.

Introducing the more convenient variable \( x_{i}=\exp (\beta J_{i}) \), where
\( \beta  \) is the inverse absolute temperature, it is straightforward to
decimate the internal degrees of freedom to obtain (Migdal-Kadanoff) recursion
relations. In this section we consider the following models:

\begin{enumerate}
\item[A.] random layered diamond lattice, Fig. \ref{fig2}(a), whose recursion relation is
\begin{equation}
\label{eq02}
x^{\prime }={R}_{\mathrm{A}}\left( x_{1},x_{2}\right) =\left( \frac{x_{1}x_{2}+q-1}{x_{1}+x_{2}+q-2}\right) ^{2};
\end{equation}

\item[B.] random branched diamond lattice, Fig. \ref{fig2}(b), with recursion relation
\begin{equation}
\label{eq03}
x^{\prime }={R}_{\mathrm{B}}(x_{1},x_{2})=\left( \frac{x^{2}_{1}+q-1}{2x_{1}+q-2}\right) \left( \frac{x^{2}_{2}+q-1}{2x_{2}+q-2}\right) ;
\end{equation}

\item[C.] random layered necklace lattice, Fig. \ref{fig2}(c), with recursion relation
\begin{equation}
\label{eq04}
x^{\prime }={R}_{\mathrm{C}}(x_{1},x_{2})=\frac{x_{1}^{2}x_{2}^{2}+q-1}{x_{1}^{2}+x_{2}^{2}+q-2};
\end{equation}

\item[D.] random branched necklace lattice, Fig. \ref{fig2}(d), with recursion relation
\begin{equation}
\label{eq05}
x^{\prime }={R}_{\mathrm{D}}(x_{1},x_{2})=\frac{x_{1}^{2}x_{2}^{2}+q-1}{2x_{1}x_{2}+q-2}.
\end{equation}

\end{enumerate}
Notice that in all these models disorder is correlated along only one spatial
direction (\( d_{1}=1 \)), while the effective dimension is \( d=2 \).
According to Eq.
(\ref{eqcrit}), we
then expect disorder to be relevant for \( \alpha >-2 \).

We now write \( x^{\prime }=x_{c}+\varepsilon ^{\prime } \) and \( x_{i}=x_{c}+\varepsilon _{i} \),
to perform Taylor series expansions about the critical point of the uniform
systems, given by \( x_{c}={R}(x_{c},x_{c}) \). For all of these models, with
\( n=2 \) independent values of the exchange parameters (along either layers
or bonds), it is straightforward to write the recursion relation
\begin{eqnarray}
\varepsilon ^{\prime } & = & w\left( \varepsilon _{1}+\varepsilon _{2}\right)
+m\left( \varepsilon _{1}^{2}+\varepsilon _{2}^{2}\right)
+f\left( \varepsilon _{1}\varepsilon _{2}^{2}
+\varepsilon _{1}^{2}\varepsilon _{2}\right) \nonumber \\
 & + & p\varepsilon _{1}\varepsilon _{2} + c\varepsilon _{1}^{2}\varepsilon _{2}^{2}
 +k\left( \varepsilon _{1}^{3}+\varepsilon _{2}^{3}\right)
 +a\left( \varepsilon _{1}^{4}+\varepsilon _{2}^{4}\right) \label{nova}
\end{eqnarray}
where \( w \), \( m \), \( p \), \( f \), \( c \), \( k \) and \( a \)
are model-dependent Taylor coefficients (that depend on the topology of the
particular models illustrated in Fig. \ref{fig2}; see Sec. \ref{criterion}).

The weak-disorder approximation \cite{Derrida and Gardner,Muzy and Salinas}
consists in assuming that
\begin{equation}
\label{wd1}
\left\langle \varepsilon \right\rangle \sim \left\langle \varepsilon ^{2}\right\rangle \sim \lambda ,
\end{equation}

\begin{equation}
\label{wd2}
\left\langle \varepsilon ^{3}\right\rangle \sim \left\langle \varepsilon ^{4}\right\rangle \sim \lambda ^{2},
\end{equation}
and in general
\begin{equation}
\left\langle \varepsilon ^{2p-1}\right\rangle \sim \left\langle \varepsilon ^{2p}\right\rangle \sim \lambda ^{p},
\end{equation}
 where \( \left\langle \cdots \right\rangle  \) is a quenched average and \( \lambda  \)
is a suitable small parameter. Within this approximation, we can use Eq. (\ref{nova})
to write recursion relations for the moments of the deviation, up to second
order in \( \lambda  \),
\begin{eqnarray}
\left\langle \varepsilon ^{\prime }\right\rangle  & = & 2w\left\langle \varepsilon \right\rangle +p\left\langle \varepsilon \right\rangle ^{2}+2m\left\langle \varepsilon ^{2}\right\rangle +2f\left\langle \varepsilon \right\rangle \left\langle \varepsilon ^{2}\right\rangle \nonumber \\
 & + & c\left\langle \varepsilon ^{2}\right\rangle ^{2}+2k\left\langle \varepsilon ^{3}\right\rangle +2a\left\langle \varepsilon ^{4}\right\rangle ,\label{eq19}
\end{eqnarray}
\begin{eqnarray}
\left\langle \left. \varepsilon ^{\prime }\right. ^{2}\right\rangle  & = & 2w^{2}\left\langle \varepsilon \right\rangle ^{2}+2w^{2}\left\langle \varepsilon ^{2}\right\rangle +4w(m+p)\left\langle \varepsilon \right\rangle \left\langle \varepsilon ^{2}\right\rangle \nonumber \\
 & + & (2m^{2}+4fw+p^{2})\left\langle \varepsilon ^{2}\right\rangle ^{2}+4wm\left\langle \varepsilon ^{3}\right\rangle \nonumber  \\
 & + & (4wk+2m^{2})\left\langle \varepsilon ^{4}\right\rangle, \label{eq20}
\end{eqnarray}
\begin{eqnarray}
\left\langle \left. \varepsilon ^{\prime }\right. ^{3}\right\rangle & = & 3w\left\langle \varepsilon \right\rangle \left\langle \varepsilon ^{2}\right\rangle +3(m+p)\left\langle \varepsilon ^{2}\right\rangle ^{2}\nonumber\\
& + & w\left\langle \varepsilon ^{3}\right\rangle +3m\left\langle \varepsilon ^{4}\right\rangle , \label{eq21}
\end{eqnarray}
 and
\begin{equation}
\label{eq22}
\left\langle \left. \varepsilon ^{\prime }\right. ^{4}\right\rangle =3w^{2}\left\langle \varepsilon ^{2}\right\rangle ^{2}+w^{2}\left\langle \varepsilon ^{4}\right\rangle .
\end{equation}

It is easy to see that there is always a non-random fixed point,
\begin{equation}
\left\langle \varepsilon \right\rangle =\left\langle \varepsilon ^{2}\right\rangle =\left\langle \varepsilon ^{3}\right\rangle =\left\langle \varepsilon ^{4}\right\rangle =0,
\end{equation}
 associated with the critical behavior of the pure model. As we pointed out
in the previous section, this fixed point becomes unstable with respect to disorder
for \( 2w^{2}>1 \). This can also be seen by an inspection of the asymptotic
behavior of Eq. (\ref{eq20}), which shows that, up to order \( \lambda  \),
the renormalized second moment
depends only on \( \left\langle \varepsilon ^{2}\right\rangle  \), with the coefficient
\( 2w^{2} \). Thus, we expect the onset of a random fixed point at a critical
value \( q_{0} \) of the number of Potts states. From the expression
\begin{equation}
\label{eq08}
x_{c}={R}(x_{c},x_{c})
\end{equation}
for the pure fixed point, we can express \( q \) as a function of \( x_{c} \)
and, using the condition \( 2w^{2}=1 \), determine the critical value \( x_{c}(q_{0}) \).
For both diamond structures displayed in Figs. \ref{fig2}(a) and \ref{fig2}(b), we have
\begin{equation}
\label{eq09}
q=\left( \sqrt{x_{c}}-1\right) \left( x_{c}-1\right) ,
\end{equation}
 and \( x_{c}(q_{0})=2.15127\ldots  \), which leads to \( q_{0}=0.53732\ldots  \).
For both necklace structures in Figs. \ref{fig2}(c) and \ref{fig2}(d), we have
\begin{equation}
q=\left( x_{c}-1\right) \left( x^{2}_{c}-1\right) ,
\end{equation}
 with \( x_{c}(q_{0})=1.46672\ldots  \), which also leads to \( q_{0}=0.53732\ldots  \).
Disorder is predicted to be relevant for \( q>q_{0} \).

We now introduce the small parameter
\begin{equation}
\lambda =x_{c}\left( q\right) -x_{c}\left( q_{0}\right) \simeq \left. \frac{dx_{c}}{dq}\right| _{q_{0}}(q-q_{0})\equiv \left. \frac{dx_{c}}{dq}\right| _{q_{0}}\Delta q,
\end{equation}
 to investigate a \( q \)-state Potts model in the immediate vicinity of the
characteristic value \( q_{0} \). It should be pointed out that, as the symmetry
of the order parameter is
one of the factors expected to determine the universality class of the models,
\( \Delta q \) is the appropriate parameter to consider. However, \( \lambda  \)
is more convenient for the algebraic manipulations. From inspection of Eqs.
(\ref{eq19})-(\ref{eq22}), we see that, up to first-order terms in \( \lambda  \),
coefficients \( w \) and \( m \) are written as
\begin{equation}
w=\frac{1}{2}\sqrt{2}+w_{1}\lambda \qquad \textrm{and}\qquad m=m_{0}+m_{1}\lambda .
\end{equation}
 It is straightforward to calculate \( w_{1}=0.13325\ldots  \), for the diamond
structures, and \( w_{1}=0.39088\ldots  \), for the necklace structures. Also,
we have \( m_{0}=-0.19088\ldots  \) and \( m_{1}=0.19865\ldots  \), for model
A; \( m_{0}=0.01849\ldots  \) and \( m_{1}=0.00758\ldots  \), for model B;
\( m_{0}=-0.48935\ldots  \) and \( m_{1}=1.22433\ldots  \), for model C; and
\( m_{0}=0.02711\ldots  \) and \( m_{1}=0.02027\ldots  \), for model D. In order to obtain
the remaining coefficients, it is enough to keep the zeroth order term in \( \lambda  \)
(see the values, up to five digits, in Table \ref{tb1}).
\begin{table}
{\centering \begin{tabular}{crrrr}
\colrule
coefficient&
model A&
model B&
model C&
model D\\
\colrule
\colrule
\( a \)&
\( -0.00926 \)&
\( 0.00917 \)&
\( -0.92623 \)&
\( 0.02894 \)\\
\colrule
\( c \)&
\( 0.08549 \)&
\( 0.00016 \)&
\( 1.38173 \)&
\( 0.07163 \)\\
\colrule
\( k \)&
\( 0.04676 \)&
\( -0.01302 \)&
\( 0.25648 \)&
\( -0.02801 \)\\
\colrule
\( f \)&
\( -0.05370 \)&
\( 0.00608 \)&                                
\( -0.33156 \)&
\( -0.04706 \)\\
\colrule
\( p \)&
\( 0.65117 \)&
\( 0.23242 \)&
\( 1.56929 \)&
\( 0.53634 \)\\
\colrule
\end{tabular}\par}

\caption{\label{tb1}Coefficients of the weak-disorder expansion for the models in Fig.
\ref{fig2}.}
\end{table}

We are finally prepared to obtain, up to lowest order in \( \Delta q \), the
non-zero values of the moments at the random fixed point. By substituting the
weak-disorder assumptions, Eqs. (\ref{wd1}) and (\ref{wd2}), into Eqs. (\ref{eq19})-(\ref{eq22}),
and then imposing consistency between equal powers of \( \Delta q \), we obtain
the leading terms for fixed values of the moments as listed in Table \ref{tb2}.
\begin{table}
{\centering \begin{tabular}{crrrr}
\colrule
moment&
model A&
model B&
model C&
model D\\
\colrule
\colrule
\( \left\langle \varepsilon \right\rangle /\Delta q \)&
\( -14.904 \)&
\( 1.0208 \)&
\( -4.4401 \)&
\( 0.34798 \)\\
\colrule
\( \left\langle \varepsilon ^{2}\right\rangle /\Delta q \)&
\( -16.170 \)&
\( -11.434 \)&
\( -1.8791 \)&
\( -2.6575 \)\\
\colrule
\( \left\langle \varepsilon ^{3}\right\rangle /(\Delta q)^{2} \)&
\( 1444.5 \)&
\( 325.73 \)&
\( 46.390 \)&
\( 39.946 \)\\
\colrule
\( \left\langle \varepsilon ^{4}\right\rangle /(\Delta q)^{2} \)&
\( 784.41 \)&
\( 392.21 \)&
\( 10.593 \)&
\( 21.187 \)\\
\colrule
\end{tabular}\par}

\caption{\label{tb2}Moments of the deviations defining the random fixed points of the
models in Fig. \ref{fig2} according to the weak-disorder expansion.}
\end{table}

In order to perform a linear stability analysis about the fixed points, we have
to calculate the eigenvalues \( \Lambda _{1} \) to \( \Lambda _{4} \) of the
matrix
\[
M_{rs}=\frac{\partial \left\langle \left. \varepsilon ^{\prime }\right. ^{r}\right\rangle }{\partial \left\langle \varepsilon ^{s}\right\rangle }.\]
As it should be anticipated from universality, it turns out that the eigenvalues (and so the
critical exponents) are the same for all models A to D. We always have two eigenvalues,
\( \Lambda _{3} \) and \( \Lambda _{4} \), whose absolute values are smaller
than unity. About the pure fixed point, we have
\begin{eqnarray}
\Lambda ^{(\mathrm{p})}_{1}&=&\sqrt{2}+0.31018\Delta q,\\
\Lambda ^{(\mathrm{p})}_{2}&=&1+0.43866\Delta q, \label{avp}
\end{eqnarray}
with a specific heat exponent
\[
\alpha _{\mathrm{p}}=-2+2.53141\Delta q.\]
At the random fixed point we have
\begin{eqnarray}
\Lambda ^{(\mathrm{r})}_{1}&=&\sqrt{2}+0.83670\Delta q,\\
\Lambda ^{(\mathrm{r})}_{2}&=&1-0.43866\Delta q,\label{avr}
\end{eqnarray}
which lead to the exponent
\begin{equation}
\alpha _{\mathrm{r}}=-2+6.82843\Delta q.
\end{equation}
From Eq. (\ref{avp}), we see that disorder becomes relevant for \( \Delta q>0 \).
Thus, as shown in Table \ref{tb2}, the weak-disorder expansion gives negative
(and thus nonphysical) values of the second moment at the random fixed point
for all models A to D. This suggests that the random fixed point in these systems
(for which \( d-d_{1}=1 \)) is nonperturbative, in agreement with numerical
calculations \cite{andelman} that predict an infinite-disorder fixed point.
Another odd feature of the weak-disorder results is that the predicted value
of the specific-heat exponent in the presence of disorder (\( \alpha _{\mathrm{r}} \))
is \emph{larger} than the corresponding quantity (\( \alpha _{\mathrm{p}} \))
for the pure model,
in disagreement with the general belief that disorder should weaken the transition.

\begin{figure}[b]
\includegraphics[width=3.5cm]{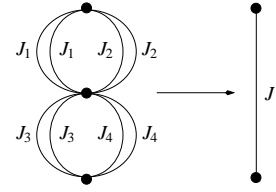}
\caption{The hierarchical lattice with $d=3$ and $d_{1}=1$ considered in Sec.
\ref{secp2}
\label{fig3}}
\end{figure}
\section{\label{secp2}A Potts model with correlated disorder:
\lowercase{$d-d_{1}>1$} case}

In order to examine the \( d-d_{1}>1 \) case, we now consider a Potts model
on a necklace hierarchical lattice \cite{andelman} shown
in Fig. \ref{fig3}, with \( d=3 \) and \( d_{1}=1 \). The unit cell contains \( n=4 \)
independent random variables and, in terms of the variables \( x_{i}\equiv \exp (\beta J_{i}) \),
the recursion relation is given by
\begin{equation}
{R}(x_{1},x_{2},x_{3},x_{4})=\frac{x_{1}x_{2}x_{3}x_{4}+q-1}{x_{1}x_{2}+
x_{3}x_{4}+q-2}.
\end{equation}

\noindent Following the same steps as in Sec. \ref{secp1}, we have
\begin{equation}
q=\left( x_{c}-1\right) \left( x^{2}_{c}-1\right) ,
\end{equation}
\( q_{0}=4+2\sqrt{2} \), and \( x_{c}(q_{0})=1+\sqrt{2} \). Performing again
the weak-disorder expansion (and truncation), and taking the average over the
disorder variables, we obtain the set of recursion relations

\begin{widetext}

\begin{eqnarray}
\left\langle \varepsilon ^{\prime }\right\rangle  & = & 4w\left\langle \varepsilon \right\rangle +2(p_{1}+2p_{2})\left\langle \varepsilon \right\rangle ^{2}+4m\left\langle \varepsilon ^{2}\right\rangle +4(f_{1}+2f_{2})\left\langle \varepsilon \right\rangle \left\langle \varepsilon ^{2}\right\rangle \nonumber \\
 & + & 2(c_{1}+2c_{2})\left\langle \varepsilon ^{2}\right\rangle ^{2}+4k\left\langle \varepsilon ^{3}\right\rangle +4a\left\langle \varepsilon ^{4}\right\rangle ,\label{eq19z}
\end{eqnarray}
\begin{eqnarray}
\left\langle \left. \varepsilon ^{\prime }\right. ^{2}\right\rangle  & = & 12w^{2}\left\langle \varepsilon \right\rangle ^{2}+4w^{2}\left\langle \varepsilon ^{2}\right\rangle +8w(3m+p_{1}+2p_{2})\left\langle \varepsilon \right\rangle \left\langle \varepsilon ^{2}\right\rangle \nonumber \\
 & + & [12m^{2}+8w(f_{1}+2f_{2})+2(p_{1}^{2}+2p^{2}_{2})]\left\langle \varepsilon ^{2}\right\rangle ^{2}\nonumber \\
 & + & 8wm\left\langle \varepsilon ^{3}\right\rangle +(8wk+4m^{2})\left\langle \varepsilon ^{4}\right\rangle, \label{eq20z}
\end{eqnarray}
\begin{equation}
\label{eq21z}
\left\langle \left. \varepsilon ^{\prime }\right. ^{3}\right\rangle =9w\left\langle \varepsilon \right\rangle \left\langle \varepsilon ^{2}\right\rangle +3(3m+p_{1}+2p_{2})\left\langle \varepsilon ^{2}\right\rangle ^{2}+w\left\langle \varepsilon ^{3}\right\rangle +3m\left\langle \varepsilon ^{4}\right\rangle ,
\end{equation}
 and
\begin{equation}
\label{eq22z}
\left\langle \left. \varepsilon ^{\prime }\right. ^{4}\right\rangle =9w^{2}\left\langle \varepsilon ^{2}\right\rangle ^{2}+w^{2}\left\langle \varepsilon ^{4}\right\rangle .
\end{equation}

\end{widetext}

\noindent It should be noted that, due to the smaller symmetry of the lattice,
we now have a larger set of coefficients. Also, notice that in this case
\( q_{0} \) is determined from the condition \( 4w^{2}=1 \).
About the critical value \( q_{0} \), and to leading order in \( \Delta q\),
we have
\begin{equation}
w=\frac{1}{2}+\frac{17\sqrt{2}-24}{4}\Delta q
\end{equation}
and
\begin{equation}
m=\frac{\sqrt{2}-2}{8}+\frac{133-94\sqrt{2}}{16}\Delta q.
\end{equation}
The values for the remaining coefficients are listed in Table \ref{tb3}.
\begin{table*}
{\centering \begin{tabular}{cccccccc}
\colrule
\( p_{1} \)&
\( p_{2} \)&
\( c_{1} \)&
\( c_{2} \)&
\( f_{1} \)&
\( f_{2} \)&
\( k \)&
\( a \)\\
\colrule
\( \frac{3\sqrt{2}}{4}-1 \)&
\( \frac{\sqrt{2}}{2}-1 \)&
\( \frac{109\sqrt{2}-144}{32} \)&
\( \frac{27\sqrt{2}-38}{32} \)&
\( \frac{25-18\sqrt{2}}{16} \)&
\( \frac{11-8\sqrt{2}}{16} \)&
\( \frac{3-2\sqrt{2}}{16} \)&
\( \frac{7\sqrt{2}-10}{64} \)\\
\colrule
\end{tabular}\par}

\caption{\label{tb3}Values of the weak-disorder coefficients for the model in Sec.
\ref{secp2}.}
\end{table*}

The moments of the deviations at the random fixed point are written as
\begin{eqnarray}
\left\langle \varepsilon \right\rangle  & = & \frac{1}{7}\left( 5-3\sqrt{2}\right) \Delta q,\nonumber \\
\left\langle \varepsilon ^{2}\right\rangle  & = & \frac{1}{7}\left( 4-\sqrt{2}\right) \Delta q,\nonumber \\
\left\langle \varepsilon ^{3}\right\rangle  & = & \frac{3}{49}\left( 95\sqrt{2}-128\right) (\Delta q)^{2},\nonumber \\
\left\langle \varepsilon ^{4}\right\rangle  & = & \frac{6}{49}\left( 9-4\sqrt{2}\right) (\Delta q)^{2}.
\end{eqnarray}

Performing a linear stability analysis about the pure fixed point we obtain
\begin{eqnarray}
\Lambda ^{(\mathrm{p})}_{1}&=&2+\left( 17\sqrt{2}-24\right) \Delta q,\\
\Lambda ^{(\mathrm{p})}_{2}&=&1+\left( 17\sqrt{2}-24\right) \Delta q,
\end{eqnarray}
with a specific-heat exponent
\begin{equation}
\alpha _{\mathrm{p}}=-1+\frac{3}{2}\frac{17\sqrt{2}-24}{\ln 2}\Delta q,
\end{equation}
while about the random fixed point we have
\begin{eqnarray}
\Lambda ^{(\mathrm{r})}_{1}&=&2-\frac{1}{7}\left( 92-65\sqrt{2}\right) \Delta q,\\
\Lambda ^{(\mathrm{r})}_{2}&=&1-\left( 17\sqrt{2}-24\right) \Delta q,
\end{eqnarray}
with
\begin{equation}
\alpha _{\mathrm{r}}=-1-\frac{3}{14}\frac{92-65\sqrt{2}}{\ln 2}\Delta q.
\end{equation}

These results show that once more disorder becomes relevant for \( \Delta q>0 \),
but now we obtain a positive (and thus physically acceptable) value of the second
moment of the deviations at the random fixed point. We also have \( \alpha _{\mathrm{r}}<\alpha _{\mathrm{p}} \).
So, as in the fully disordered model (\( d_{1}=0 \)) studied by Derrida and
Gardner \cite{Derrida and Gardner}, and in agreement with numerical calculations
\cite{andelman}, the weak-disorder scheme predicts a (perturbative) finite-disorder
fixed point, with values of the critical exponents continuously approaching
those of the pure model as \( \Delta q\rightarrow 0 \).

\section{An Ising model with correlated disorder}
\label{secising}
\begin{figure}
\includegraphics[width=3.5cm]{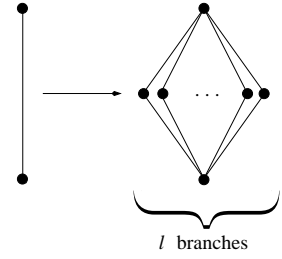}
\caption{A diamond hierarchical lattice with $b=2$ bonds and $l$ branches.
\label{fig4}}
\end{figure}

The set of recursion relations given by equations (\ref{eq19}) to (\ref{eq22}),
with a suitable redefinition of parameters, can also be used to analyze an Ising
model on a more general diamond structure with \( b=2 \) bonds and \( l \)
branches, and correlated disordered ferromagnetic exchange interactions along
the layers (see Fig. \ref{fig4}). For this structure we also have \( d-d_{1}=1 \). While
in the Potts models we have a natural parameter, \( q \), for varying \( \alpha  \),
we now change the topology of the lattice, by varying \( l \), to obtain the
same effect.

Using the standard Ising Hamiltonian,
\begin{equation}
\label{eq1}
{\mathcal{H}}_{I}=-\sum _{(i,j)}J_{i,j}\sigma _{i}\sigma _{j},
\end{equation}
 with \( \sigma _{i}=\pm 1 \), and introducing the more convenient
transmissivity variable
\( t_{i}=\tanh \beta J_{i} \), the decimation of the intermediate spins leads
to the recursion relation
\begin{equation}
\label{eq3}
t^{\prime }={R}_{l}(t_{1},t_{2})=\tanh \{l\tanh ^{-1}(t_{1}t_{2})\}.
\end{equation}
 As in Sec. \ref{secp1}, we now write \( t^{\prime }=t_{c}+\varepsilon  \)
and \( t_{i}=t_{c}+\varepsilon _{i} \), where
\begin{equation}
\label{311}
t_{c}={R}_{l}(t_{c},t_{c})
\end{equation}
 is the critical transmissivity of the uniform model. We then perform quenched
averages, and use the weak-disorder assumption, to obtain Eqs. (\ref{eq19})
to (\ref{eq22}).

The critical parameters for relevance of disorder, \( l_{0}=1.44976\ldots  \)
and \( t_{c}(l_{0})=0.79951\ldots  \), come from Eqs. (\ref{311}) and (\ref{eq11}).
The small parameter \( \lambda  \) can be chosen as
\begin{equation}
\lambda =t_{c}\left( l\right) -t_{c}\left( l_{0}\right) =\left. \frac{dx_{c}}{dl}\right| _{l_{0}}(l-l_{0})\equiv \left. \frac{dx_{c}}{dl}\right| _{l_{0}}\Delta l.
\end{equation}
Again we use \( \lambda  \) as a convenient parameter for algebraic manipulations,
although \( \Delta l \) is the physically relevant variable. The Taylor coefficients
in Eqs. (\ref{eq19}) to (\ref{eq22}) are given by \( w=\sqrt{2}/2-0.54522\lambda  \),
\( m=-0.49698-0.65422\lambda  \), \( a=0.11520 \), \( c=-1.64903 \), \( k=-0.12543 \),
\( f=-1.61924 \), and \( p=-0.10953 \). We then calculate the leading values of the
moments at the random fixed point,
\begin{eqnarray}
\left\langle \varepsilon \right\rangle  & = & -0.64971\Delta l,\nonumber \\
\left\langle \varepsilon ^{2}\right\rangle  & = & -0.27076\Delta l,\nonumber \\
\left\langle \varepsilon ^{3}\right\rangle  & = & -0.30084(\Delta l)^{2},\nonumber \\
\left\langle \varepsilon ^{4}\right\rangle  & = & +0.21993(\Delta l)^{2}.
\end{eqnarray}
A linear stability analysis leads to the eigenvalues \( \Lambda _{1}^{(\mathrm{p})}=\sqrt{2}+0.71884\Delta l \)
and \( \Lambda _{2}^{(\mathrm{p})}=1+1.01659\Delta l \), for the pure fixed
point, and \( \Lambda _{1}^{(\mathrm{r})}=\sqrt{2}+1.20537\Delta l \) and \( \Lambda _{2}^{(\mathrm{r})}=1-1.01659\Delta l \),
for the random fixed point. From these values, we see that disorder is relevant
for \( \Delta l>0 \), but we again have \( \left\langle \varepsilon ^{2}\right\rangle <0 \)
in this case.

We then obtain the specific heat critical exponents
\begin{equation}
\alpha _{\mathrm{p}}=-1.07163+2.51471\Delta l
\end{equation}
and
\begin{equation}
\alpha _{\mathrm{r}}=-1.07163+5.56379\Delta l.
\end{equation}
 For \( \Delta l<0 \), which corresponds to \( \alpha <-1.07163\ldots  \),
the pure fixed point is stable and the random model displays the same critical
behavior as its pure counterpart. For \( \Delta l>0 \), which corresponds to
\( \alpha >-1.0713\ldots  \) (yielding again \( \alpha _{\mathrm{r}}>\alpha _{\mathrm{p}} \)),
we anticipate a novel class of (random) critical behavior, but the fixed
point must be nonperturbative, as suggested by the nonphysical character of
the weak-disorder results.

\section{Conclusions}

We have used a weak-disorder scheme and real-space renormalization-group
techniques to look at the effects of correlated disorder on the critical
behavior of some $q$-state Potts models with correlated disordered
ferromagnetic interactions along $d_{1}$ out of $d$ spatial dimensions. We
have written exact recursion relations on diamond and necklace hierarchical
structures, which are equivalent to Migdal-Kadanoff approximations for the
corresponding Bravais lattices.

The weak-disorder scheme leads to analytical results by truncating the
recursion relations for the moments of the distribution function. We first
used scaling arguments to rederive a general expression for the Harris
criterion to gauge the relevance of disorder (and show that it is related to
the number of independent random variables in the unit cell of the lattice
and the first derivative of the recursion relations at the pure fixed
point). We then performed a number of calculations to compare with numerical
findings by Andelman and Aharony.

For $q$-state Potts models on various hierarchical lattices with
ferromagnetic random exchange interactions correlated along $d_{1}=1$ out of
$d=2$ dimensions, we obtained a new (random) fixed point for $q$ larger than
a characteristic value $q_{0}$, where disorder becomes relevant. This fixed
point, however, is located in a nonphysical region of parameter space, which
suggests that a nonperturbative (infinite-disorder) fixed point must be
present (as pointed out by the calculations of Andelman and Aharony). For a $%
q$-state Potts model on a diamond lattice with $d_{1}=1$ and $d=3$, we
obtained a physically acceptable, finite-disorder fixed point, for $q>q_{0}$%
, as in the fully disordered model analyzed by Derrida and Gardner (although
in our case the usual expression of the Harris criterion is not fulfilled).
Also, we considered an Ising model ($q=2$) on a diamond lattice with $b=2$
bonds and $l$ branches (where $l$, instead of $q$, is the control
parameter), which is another example of a $d-d_{1}=1$ system. Again, the
weak-disorder expansion predicts a nonphysical random fixed point.

To summarize the results of this paper, we point out that, in the vicinity of
the point where disorder becomes relevant,
the weak-disorder scheme always produces a perturbative random fixed point,
but there are two distinct possibilities, depending on the difference between
$d$ and $d_{1}$:
(i) If $d-d_{1}=1$, the perturbative fixed point is characterized by a negative
variance, and is thus nonphysical, suggesting the existence of another,
nonperturbative fixed point; (ii) If $d-d_{1}>1$, the scheme predicts a
physically acceptable perturbative fixed point. It should be mentioned that
this same picture holds for fairly general hierarchical
lattices, in particular those with noniterating bonds, as considered by Griffiths
and Kauffman \cite{Grif}. Furthermore, in the case of the quantum Ising model
with bond disorder,
which corresponds to the extreme-anisotropy limit of the two-dimensional
McCoy-Wu model ($d-d_{1}=1$), Fisher \cite{Fisher} was able to obtain a
(presumably exact) fixed-point probability distribution with infinite variance.
It is certainly interesting to investigate whether similar
conclusions still hold for other models (as the problem of directed polymers
in random environments \cite{Muk}) on either hierarchical or Bravais
lattices.

\begin{acknowledgments}
This work was partially financed by
the Brazilian agencies
CNPq and Fapesp.
\end{acknowledgments}


\end{document}